\documentclass[12pt,preprint]{aastex}

\usepackage{amsfonts}
\usepackage{amsmath}
\usepackage{amssymb}
\usepackage{pdflscape}
\usepackage{rotating}
\usepackage{natbib}
\usepackage{txfonts}
\usepackage{longtable}
\usepackage[small,bf]{subfigure}
\usepackage{multirow}
\usepackage{xspace}
\usepackage[pdftitle={AGILE-SAM-Bulgarelli-etal.},colorlinks=true, citecolor=blue, linkcolor=blue, urlcolor=blue]{hyperref}

\def \aa {A$\&$A$\;$}

\def \sas {\textit{IASFBO SAS pipeline }}
\def \qls {\textit{ADC QLS pipeline }}
\def \sasnospace {\textit{IASFBO SAS pipeline}}
\def \qlsnospace {\textit{ADC QLS pipeline}}
\def \degmark{^\circ}

\def \phcmsec{\hbox{photons cm$^{-2}$ s$^{-1}$}}
\def \cmsecsr{\hbox{cm$^{-2}$ s sr}}

\def \gray {$\gamma$-ray }
\def \enmev {$E > 100$  MeV }

\def \mtt {  }

\def \abb {   }

\def \mmt {   }

\def \abp { }

\shorttitle{The AGILE Alert System for Gamma-Ray Transients}
\shortauthors{Bulgarelli et al.}

\begin{document}


\title{THE AGILE ALERT SYSTEM FOR GAMMA-RAY TRANSIENTS}


\author{A.~Bulgarelli\altaffilmark{1}, M.~Trifoglio\altaffilmark{1}, F.~Gianotti\altaffilmark{1},
M.~Tavani\altaffilmark{2,3}, N.~Parmiggiani\altaffilmark{4},
V.~Fioretti\altaffilmark{1}, A.W.~Chen\altaffilmark{5,6}, S.~Vercellone\altaffilmark{7},
C.~Pittori\altaffilmark{8}, F.~Verrecchia\altaffilmark{8},
F.~Lucarelli\altaffilmark{8}, P.~Santolamazza\altaffilmark{8},
G.~Fanari\altaffilmark{8}, P.~Giommi\altaffilmark{8},
D.~Beneventano\altaffilmark{4}, A.~Argan\altaffilmark{2},
A.~Trois\altaffilmark{2},  E.~Scalise\altaffilmark{2},  F.~Longo\altaffilmark{9}, A.~Pellizzoni\altaffilmark{10}, G.~Pucella\altaffilmark{11},  S.~Colafrancesco\altaffilmark{12},
V.~Conforti\altaffilmark{1}, P.~Tempesta\altaffilmark{13},
M.~Cerone\altaffilmark{13}, P.~Sabatini\altaffilmark{14},
G.~Annoni\altaffilmark{14}, G.~Valentini\altaffilmark{15}, and
L.~Salotti\altaffilmark{15} }

\altaffiltext{1}{ INAF/IASF--Bologna, Via Gobetti 101, I-40129 Bologna, Italy }
\altaffiltext{2}{ INAF/IASF--Roma, Via del Fosso del Cavaliere 100, I-00133 Roma, Italy }
\altaffiltext{3}{ Dip. di Fisica, Univ. ``Tor Vergata'', Via della Ricerca Scientifica 1, I-00133 Roma, Italy }
\altaffiltext{4}{ University of Modena and Reggio Emilia, Dipartimento di Science e Metodi dell'Ingegneria, Italy }
\altaffiltext{5}{ INAF/IASF--Milano, Via E.~Bassini 15, I-20133 Milano, Italy }
\altaffiltext{6}{ School of Physics, University of the Witwatersrand, Johannesburg Wits 2050, South Africa}
\altaffiltext{7}{ INAF/IASF--Palermo,  Via U. La Malfa 153, I-90146 Palermo, Italy }
\altaffiltext{8}{ ASI--ASDC, Via G. Galilei, I-00044 Frascati (Roma), Italy }
\altaffiltext{9}{ INFN Trieste, I-34127 Trieste, Italy}
\altaffiltext{10}{ INAF-OAC, I-09012 Capoterra, Italy}
\altaffiltext{11}{ ENEA Frascati, I-00044 Frascati (Roma), Italy}
\altaffiltext{12}{ INAF-OAR, I-00040 Monteporzio Catone, Italy}
\altaffiltext{13}{ Telespazio, Fucino Space Centre ÒPiero FantiÓ, Via Cintarella 67050, Ortucchio (Aq) - Italy }
\altaffiltext{14}{ CGS S.p.A, Via Gallarate 150, Milan, Italy }
\altaffiltext{15}{ ASI, Via del Politecnico snc , I-00133 Roma, Italy}


\begin{abstract}

In recent years, a new generation of space missions offered great opportunities of discovery in high-energy astrophysics. In this \abp{article} we focus on the scientific operations of the Gamma-Ray Imaging Detector (GRID) \abp{onboard} the \textit{AGILE} space mission. The \textit{AGILE}-GRID, sensitive in the energy range \abp{of} 30 MeV$-$30 GeV, has detected many \gray transients of \abp{galactic and extragalactic origins}. This work presents the \textit{AGILE} innovative approach to fast \abp{\gray transient} detection, \abp{which is} a challenging task and a crucial part of the \textit{AGILE} scientific program. The goals are to describe: (1) the AGILE Gamma-Ray Alert System, (2) a new algorithm for blind search identification of transients within a short processing time, (3) the \textit{AGILE} procedure for \gray transient alert management, and (4) the likelihood \abp{of ratio tests that} are necessary to evaluate the \abp{post-trial} statistical significance of \abp{the} results. Special algorithms and an optimized sequence of tasks are necessary to reach our goal. Data are automatically analyzed at every orbital downlink by an alert pipeline operating on different timescales. As proper flux thresholds are exceeded, alerts are automatically generated and \abp{sent as} SMS messages to cellular telephones, e-mails, and push \abp{notifications} of an application for smartphones and tablets. These alerts are crosschecked with the results of two pipelines, and a manual analysis is performed. Being a \abp{small scientific-class mission}, \textit{AGILE} is characterized by optimization of both scientific analysis and \abp{ground-segment} resources. The system is capable of generating alerts within \abp{two to three hours of a} data downlink, an unprecedented reaction time in \gray astrophysics. 

\end{abstract}





\section{Introduction}
\label{sec:intro}

Gamma-ray astrophysics above 100 MeV from space \abp{have} substantially
progressed in recent years. After a first pioneering phase, space
observatories
such as \textit{COS-B} (\abp{from 1975 to 1982;}
\cite{mayer, swanenburg}) and
the \textit{Compton Gamma-Ray Observatory} (\textit{CGRO}; \abp{from 1991 to 2000;}    \cite{fichtel}) were able to 
successfully \abp{map} \gray sources and extended Galactic
regions. The field of view (FoV) of the \gray instrument EGRET 
\abp{onboard of \textit{CGRO} and} operating in the 100 MeV$-$10 GeV domain was
sufficiently large \abp{enough} (0.5 sr) to simultaneously monitor many sources
in the Galactic plane and at high Galactic latitudes
\cite{fichtel,thompson}. The EGRET point-spread function (PSF;
6$-$7$\degmark$ for the 68\% containment radius at 100 MeV)
allowed a first mapping of the Galactic plane and source
positioning. Today, with the advent of silicon-based \gray
detectors \textit{AGILE} \cite{Tavani_2009} and
\textit{Fermi}-LAT \cite{Atwood_2009}, both the PSF and FoV \abp{have}
substantially improved \abp{since} the 1990's. For both \textit{AGILE}
and \textit{Fermi}-LAT, the FoV is now quite large (about
2.5 sr), covering more than 1/5 of the whole sky, and the PSF at
100 MeV has a containment radius of 3$-$4$\degmark$
\cite{abdo2,Chen_2013}. Large portions of the sky can then be
simultaneously monitored with optimal \abp{source detection capabilities}. The issue of optimizing source monitoring and the
search for \gray transients is \abp{thus} of great
importance.

\textit{AGILE} (\textit{Astrorivelatore Gamma ad Immagini LEggero $-$ Light Imager
for Gamma-ray Astrophysics}) is a scientific mission of the
Italian Space Agency (ASI) \abp{that was launched on 2007  April 23}
\cite{Tavani_2009}. The Gamma-Ray Imaging Detector (GRID) \abp{is used} for
observations in the 30 MeV$-$30 GeV energy range.  The \textit{AGILE}
orbital characteristics (quasi-equatorial with an inclination
angle of 2$\degmark$.5 and \abp{an average altitude of} 530  km,
$96^\prime$ period) are optimal for low-background \gray
observations.
\abp{From 2007 July to 2009 October,} \textit{AGILE} observed the
\gray sky in the so-called ``pointing mode", \abp{which is} characterized by
quasi-fixed pointings with a  slow drift ($\sim 1\degmark$/day)
of the instrument boresight direction following solar panel
constraints.
Due to a change \abp{in} the satellite pointing system control,
since \abp{2009 November} the \textit{AGILE} \gray observations
\abp{have been} obtained  with the instrument operating in
``spinning mode" (\abp{i.e.,} the satellite axis sweeps a 360$\degmark$ circle
in the sky every approximately $7^\prime$). Depending on the
season, the whole sky is progressively exposed.

 The variability aspect  of
the \gray sources is a key factor for our understanding of their
origin and nature \cite{ackermann,Hinton_2013}, and the study of the variability of the \gray sky above 30 MeV
requires   special focusing and resources. The task can be
daunting: a large fraction of currently known \gray sources
are unidentified\footnote{\abp{containing about one-third of the sources found in the current catalogues of EGRET} \cite{Hartman_1999}, \textit{AGILE}
\cite{Pittori_2009}  and \textit{Fermi} \cite{Nolan_2012} \abp{have
no known or confirmed counterparts at other wavelengths.}}, and in general
many sources show statistically significant variability at all
timescales (hours, days, weeks).

The \gray transient monitoring program is a very
important task of the \textit{AGILE} science program. It \abp{became}
active in a preliminary fashion at the beginning of the mission
(\abp{the } summer of 2007) and it has been improving ever since.
 It is a dedicated alert system \abp{that} is implemented within
 the AGILE Ground Segment. \abp{Because
detecting time variable sources during their early phases of emission and fast communication
with the astrophysics community are crucial, the AGILE Alert SystemÕs} goal is  to provide
 excellent quality \gray data \abp{within} the shortest time \abp{possible} for scientific
 analysis and follow-up observations. This task requires a necessarily automated detection
system \abp{that} can
issue alerts immediately upon detection of \gray flares, enabling
then the study of the transient sky. In this \abp{article},
we focus on the innovative features of the scientific analysis
performed at the \textit{AGILE} ground segment level. In particular, we
 describe the successful strategy for fast  analysis of
\gray data, a challenging task that requires a special optimization of  resources (Section \ref{S2}). We \abp{also} discuss the role played by special
 software tools and algorithms  (Section \ref{S3}).
Section \ref{S4} presents the team organization  and   the tools developed for mobile devices to support the manual
procedures to confirm automatic alerts. Section \ref{S5} focus on the
characterization of the statistical significance of pipeline alerts.
 Section \ref{S6} summarizes some of the scientific results
 provided to the astrophysical community; we discuss, as a concrete example,
  the reaction time for  the very important case of the 2010 Crab Nebula \gray flare discovered by \textit{AGILE}.  

The main goals of this
\abp{article} are \abp{to describe}: (1)  the AGILE
Alert System, \abp{focusing} on the pipeline developed at
IASFBO;  (2)  a new algorithm
(``spotfinder") for blind search identification of flare candidates
within a short processing time;  (3)  the \textit{AGILE}
procedure for alert management; and (4) an evaluation of
the statistical distribution of the data products in
order to properly evaluate the false detection occurrence rate.

\section{General overview}
\label{S2}

This section describe the general architecture of the
AGILE Ground Segment from data acquisition to alert
generation (see Figure \ref{fig_F1}).

\textit{AGILE} data are downlinked to the ASI Malindi ground
station in Kenya every orbit (about every 96 minutes
\abp{which is} the time window for data retrieval, and commands transmitting is
10 minutes
per orbit). Data are then quickly
transferred via ASINet through a dedicated Intelsat
\abp{bidirectional} link to Telespazio, Fucino (AQ, Italy), where the
Mission Operation Center (MOC) is located. Data are then
transferred  to the ADC, \abp{which is} located in Frascati (Italy), for data
reduction, archiving, and distribution.

Raw data are routinely archived at the ADC, and then
converted \abp{to} FITS format L1 (\abp{preprocessed} data) through the \textit{AGILE}
\abp{Preprocessing} System \cite{Trifoglio_2008}. Data are
further processed (L2, or event list and L3, or sky maps) using
software tasks developed by the \textit{AGILE} instrument team  and
then integrated into the pipeline systems developed at ADC
for quicklook monitoring and consolidated archive generation.
The L0 and L1 data are  forwarded to the IASFBO site
\cite{Trifoglio_2007} where the \sas runs\footnote{\abp{there is also a backup chain that runs from Telespazio to Compagnia Generale dello
Spazio (CGS, Milan) and then to the IASFBO site that is activated only when problems
occur in the nominal flow chain and only after an authorization is received from the Mission
Director.}}.

The AGILE  Alert System
 is composed of two independent automated \abp{analyses} parts:
 (1) the \textit{Quick-Look Scientific (QLS) pipeline}
 running at the \textit{AGILE Data Center (ADC)}\footnote{ http://agile.asdc.asi.it}
 (hereafter \qlsnospace), \abp{which is} part of the ASI Science Data Center
 (ASDC); \abp{and the (2)}
 \textit{AGILE-GRID Science Alert System (SAS) pipeline}  running
 at the  \textit{INAF/IASF Bologna (IASFBO)}\footnote{ http://gtb.iasfbo.inaf.it}
 (hereafter \sasnospace).
  These systems process the data with different levels of data
  processing and procedures \abp{while} looking for transient \gray events.
  Both systems operate in \abp{real-time} mode, with
automatic processing starting as soon as data
are available. The \sas generates an alert to the AGILE
Team for each candidate \gray flare within \abp{two to three} hours from the time
of the last
orbit data acquisition stored in the satellite. In doing so, preference is given to
speed rather than completeness of orbital telemetry.
 On the other hand,   \qls generates a daily report containing a list of candidate flares,
 sent via \abp{e-mail} using a more complete data archive.

 The final consolidated archive of the \textit{AGILE} Mission is then produced at
 ADC by a dedicated
 pipeline\footnote{\abp{more details on the ADC organization and functionalities
 will be published in \abp{C. Pittori et al. (2014, in preparation)}. See also \cite{Pittori_2013}.}}.

As  data of the latest acquired orbit are received
at the IASFBO site, 
data processing 
by the \sas 
starts immediately running on \abp{a} cluster system of 64 CPUs.
Processing time typically lasts
about $65^\prime$$-$$85^\prime$ \abp{in order} to analyze in parallel all accessible
sky regions,
taking into account  the \gray exposure.
In order to detect transients or variable sources
the \textit{Flare Search Procedure  1} \abp{subsystem} performs an
automated search for transient \gray emission. The output is
then analyzed 
to select the best transient candidates.
Details on the time duration required to perform
each step of data transfer and elaboration are reported in Figure
\ref{fig_time}.

Another independent subsystem, (the \textit{Flare Search Procedure
2}), is part of the \qlsnospace. 
The procedure uses a more complete data set to consolidate the
data of the current orbit. The procedure is necessarily
slower (it requires about 3.5 \abp{to 5} hours after data acquisition), \abp{as}  it is necessary to wait the following orbit to
generate an output, because the data to reconstruct the last minutes  of the satellite pointing direction are contained in the next orbit).
The ADC result is in the form of a daily report \abp{that}
lists all candidate \gray transients, and \abp{it} is distributed
via e-mail to the team twice a day.

When the AGILE Team receives a high confidence automatic
alert from the \sasnospace, 
a more refined analysis is undertaken
before communication to the astronomical community.
A visual check of counts and intensity maps is performed via \abp{a} web
browser and  also through
 smartphones. The alert is analyzed in parallel with \abp{the online data archives of both} ADC and IASFBO: the IASFBO site acts also
as a point of distribution to the AGILE Team Center (ATC) (located
at INAF/IAPS, Rome) of the L2 and L3  data generated by
\textit{SAS} itself. This L2/L3 operative archive is  used in the
daily quick-look monitoring activities.

A dedicated  Application (App)  for smartphones and tablets (the
\textbf{AGILEScience
App}\footnote{https://itunes.apple.com/it/app/agilescience/id587328264?l=en$\&$mt=8 and https://itunes.apple.com/it/app/agilescience-for-ipad/id690462286?l=en$\&$mt=8})
is also connected with the \sas  and has a reserved area where the
scientific results are made visible to the AGILE Team.
Processed results can also be made available to the public through
this App (see Section \ref{SAPP}).


\section{The IASFBO SAS Analysis Procedure}
\label{S3}

The \textit{Flare Search Procedure   1} of the \sas 
\abp{automatically performs} a search for transient \gray emissions in
two steps:
\begin{enumerate}
\item \abp{a} selection of an ensemble of models of 
transient sources  (i.e., a selection of a list of
candidate \gray flaring sources ) with two independent methods:
 a blind search procedure called \textit{spotfinder}  (see
Section \ref{sec32}) and  \abp{the} use of a list of known
sources from
selected astrophysical source catalogues; \abp{and}

 \item  calculation of the  flux and
significance level for each \gray
transient candidate (see Section \ref{sec33}) .
\end{enumerate}

This procedure produces a list of candidate flares with
its associated statistical significance. Alerts are
generated for a subclass of these selected candidates 
(see Section \ref{sec34}).

Data are analyzed for each orbit, producing a sliding
window in the generated light-curves of \gray sources. An example
is reported in Figure \ref{fig_crab2}.

\subsection{Data Analysis Method}
\label{sec31}

The \textit{AGILE}-GRID data analysis is currently performed on the data
set generated with the software package (version 4) publicly
available at the
 ASDC  \abp{website and uses} the FM3.119 background event filter \abp{(see A. Bulgarelli et al. (2014, in preparation))} and the I0023 version
 calibration matrices \citep{Chen_2013}. The events collected during
 the passage in the South Atlantic
 Anomaly and  Earth albedo background photons are removed from the pipeline.
 To reduce  particle background contamination, we select only events
 flagged as confirmed \gray events (\abp{G-class} events, corresponding to a
 sensitive area of $\sim$ 330 cm$^2$ at 100 MeV). \textit{AGILE} counts, exposure,
 and Galactic diffuse emission background maps are generated with a bin size
 of $0\degmark.3 \times 0\degmark.3$ to compute the period-averaged source
 flux and its evolution.

 Owing to the relatively low event detection rate and the
 extent of the \textit{AGILE}-GRID PSF, a binned \abp{multisource
 Maximum Likelihood Estimator} (MLE) developed at INAF/IASF Milano is used to search for transient
 emission from \gray sources. This method iteratively optimizes \abp{the} position
 and flux of all the sources of the region \cite{Bulgarelli_2012}. The use of a binned likelihood method greatly increases the speed of analysis with respect to an un-binned version. The possibility to optimize at the same time source flux and position  has simplified the development of the pipeline.  The same tool is used also in the manual verification procedure (see Section \ref{S4}).  

  The likelihood ratio test $T_s$ calculated by the MLE
  is used to compare two ensembles of hypotheses,
 one in which a \gray source is present and another
 (the null hypothesis) in which it is absent. 
 More specifically,
 the null hypothesis can be formulated as an ensemble
 of models \abp{that keep} the flux of the flaring source fixed to zero
 and the fluxes of steady sources fixed to their known fluxes (if any).
 For the alternative hypothesis (a flaring source is present),
 the flux and position of this source
 are allowed to be free, and the fluxes of steady sources
 are fixed to their known fluxes. 
 The analysis is performed over a region of $10\degmark$ radius.
 The Galactic diffuse \gray radiation \cite{Giuliani_2004}
 and the isotropic emission are taken into account in the model with two parameters:
 (1) $g_{gal}$, the coefficient of the Galactic diffuse emission model, and
 (2) $g_{iso}$, the isotropic diffuse intensity.  As noted in \citep{Bulgarelli_2012}  the isotropic component is dominated by instrumental charged particle background rather than by the extragalactic diffuse emission, in contrast to data from EGRET and \textit{Fermi}-LAT. This instrumental charged particle changes over time and space and for these reasons they are kept free during the data analysis.

We analyze the sky within the current \textit{AGILE} FoV over one-day
intervals during the ``pointing mode" period and over two-day
intervals during the ``spinning mode" period 
\abp{to obtain approximately similar exposures} for both ``pointing" and ``spinning" at
the center of the exposed region.
We identify the transient episodes, analyzing both the
100 MeV$-$30 GeV and 400 MeV$-$30 GeV energy bands.

\subsection{Automated Blind Search for Transient $\gamma$-Ray Emissions}
\label{sec32}

The peculiarity of the \textit{AGILE}-GRID instrument has required the development of a new algorithm called \textit{spotfinder} that extracts ``candidate \gray flares" from Gaussian smoothed  \gray sky maps  (either counts or intensity maps are used); the automated blind search for transient \gray emission procedure is based on this algorithm, that takes into account the following issues:
\begin{enumerate}
\item  The FoV of the \textit{AGILE}-GRID instrument  usually contains both the Galactic plane and the extra-Galactic sky, that are very different in terms of
\begin{enumerate}
\item background components (the Galactic plane is dominated by the diffuse \gray emission, the extra-Galactic sky is dominated by the isotropic emission);
\item the number of \gray sources in the same area of the sky.
\end{enumerate}
\item the \textit{AGILE}-GRID instrument has a PSF of 2$\degmark$.1 at 30$\degmark$ off-axis for $E > 100$ MeV reconstructed energy (and spectral index $\alpha = -1.66$ from Monte Carlo data). This algorithm takes into account this extension;
\item due to the fact that the  isotropic component is dominated by the charged particle background that may change over
space,  the \textit{spotfinder} algorithm estimates the background
components (diffuse, isotropic and instrumental) for each
sub-region of the \gray sky map.
\end{enumerate}
The development and optimization
of this
algorithm was one of the main tasks for the development of our
automated data analysis system for detection of \gray flares. 
\textit{Spotfinder} is basically a connected component labeling algorithm \citep{Rosenfield_1966}
 that works with \abp{multilevel} images.
Each \gray emission  can be viewed as a connected component region.
 It uses a labeling algorithm \cite{DiStefano_1999} that searches \abp{for}
 the connected regions \abp{in} a binary image
 (an image with two colors, black and white).
 
Each counts or intensity map is segmented into three different
regions ($b < -10\degmark$, $| b | < 10\degmark$, \abp{and} $b > 10\degmark$) to account for the
higher background and source confusion near the Galactic plane
(see first panel of Figure \ref{fig_spotfinder}). After the
selection of a sky region, there are some options that can change
the value of the pixels: (1) if the value of a pixel is above a
\abp{well-defined} threshold, it is set to zero to maximize the
effectiveness of the search in regions with a low \gray emission;
(2) if the value of the exposure is below a threshold, the
related pixel in the counts or intensity maps is set to zero to
avoid searches in sky regions with a low level of exposure; and (3)
it is possible to  subtract two maps. In the current version of
the algorithm used in the \sasnospace,  only  option (2) is used with
a typical value of 50 $\cmsecsr$ for maps with a bin size of
0$\degmark$.3.

A Gaussian smoothing of the map is performed with a typical kernel
of \abp{three} bins. In a \gray smoothed map the values of a pixel range
from 0 to a maximum value within a continuous interval of values.
After  smoothing, the intervals are normalized and discretized to
$N$ intervals, \abp{which rounds} each value to the nearest integer value
(see \abp{the} second panel of Figure \ref{fig_spotfinder}) \abp{and divides} the
original image  into $N$ images ($N$=100 is a typical value)
containing only pixels with the same value $k$ (see \abp{the} third panel of
Figure \ref{fig_spotfinder}  with $N=8$).  The connected component
search procedure starts from these normalized and
discretized maps.

The search for connected component regions is an iterative
procedure. The procedure starts by considering only the image that
contains the pixels with value $N$  and then \abp{by} calculating
the connected component regions contained into this first image.
For each iteration $k$ ($k=N-1, $...$, 1$) the level $k-1$ is
added to the current image, and new connected component regions
are calculated. The effect of merging two levels is that the
original regions grow by adding the pixels of the neighboring
level. This growing procedure stops when more than $M$ connected
regions are found (where $M$ is a parameter, a typical value is
8). At the end, for each connected region the barycenter is
calculated. This is the starting position for the MLE  (see an example
of the found connected regions in the
  first panel of Figure \ref{fig_spotfinder}).

There are also some additional options \abp{that are} applicable after the
determination of the connected regions: (1) if two connected
regions are too close,
 only the first connected region survives; \abp{and}
(2) if a connected region is outside a circle centered in the
center of the map with a radius specified by the user, this region
is discarded.

\subsection{Calculation of the Significance of Each Candidate Flare}
\label{sec33} The second step of the automated search procedure
for transient \gray emission is the calculation of the confidence
level of each candidate flaring source, which is
 performed in two \abp{substeps} using the MLE. In the first \abp{substep},
 all sources included in the initial ensemble of models
 (ordered using the intensity value of the bins contained
 in the connected region and above a \abp{predefined} exposure level
 threshold)
 are analyzed;
 the flux of the candidate flares is allowed to vary
 and the position is kept fixed at the value of the
 center of the connected region. This step is useful
 to reduce the number of candidates for the final evaluation,
 \abp{which minimizes}
the complexity of the model. In the final step,  only detections
with $\sqrt{T_s} \ge 2$ are selected and \abp{reanalyzed}
simultaneously. The flux and position of the candidate flares are
allowed to vary, and the spectrum of each candidate
 is assumed to be a \abp{power law} with the spectral index kept fixed to  2.1.
In the end,  we obtain a list of candidate transient sources
and their \abp{pretrial}
statistical significance.

\subsection{Science Alert Generation}
\label{sec34}

For each candidate transient event with $\sqrt{T_s} > 25$
(corresponding to $\sigma \ge 4.3$; see Section
\ref{S5}) an alert is generated and sent via SMS, e-mail and
through the notification system of the
 iOs \textit{AGILEScience App} (see Section \ref{SAPP}).

The generated \abp{e-mail} contains the following information:
\begin{itemize}
\item counts, flux, and related errors; 
\item the optimized
position of the source, error circle, and ellipse at $95\%$
confidence \abp{levels};
\item a list of possible associations within the
error circle based on a list of known sources; 
\item a short
description of the parameters of the analysis; 
\item the exposure
value; 
\item some useful links to NED and SIMBAD to check the
presence of known sources within the error circle; 
\item a link to
the web directory containing the analyzed data and the overall
results; and
\item the detailed result of the run
resulting in a detection (the overall list of sources and
the background estimation parameters are provided).
\end{itemize}

The SMS and the App \abp{notifications contain} a reduced version of the
\abp{e-mail} content. These notifications \abp{only} include
 the significance level, the flux and related error, the
exposure level, a possible association, and the list of 
analysis parameters.

\section{The Final Verification  Procedure}
\label{SMAN}
\label{S4}

For the most interesting automatic detections, a
final verification is performed by human
intervention.
During routine daily monitoring, two scientists on
duty are assigned to check the alerts generated by the \sasnospace, one
from the AGILE Team (at INAF IASF Bologna, Milan, Palermo, IAPS
Rome, and INFN Trieste) and one at \abp{the} ASDC.
Alerts automatically generated by the \sas are
cross-checked with the data and daily reports generated by the
\qlsnospace.  \abp{The} alerts generated by the IASFBO pipeline are
received about two hours before the high quality L2 data generated
by the ASDC pipeline, \abp{and} this enables the AGILE Team to gain time
during the
 final check.

This is a key point of the overall AGILE Alert System.
We can perform a 
check before the refined version of
the scientific data is available. We can then
optimize 
the results of this analysis  with the IASFBO raw data
archive, and check the results with \abp{the} ADC consolidated archive when
available. Both pipelines work with the common goal \abp{of producing}
scientific results in the shortest possible time and with the best
data quality.
Only detections of \abp{pretrial} significance larger than $5
\sigma$ (see  Section \ref{S5}) that survive this 
 check by human intervention
 are candidates for fast alerts to be distributed to the
 community.
The significance threshold 
might occasionally be lower if there is independent
evidence of simultaneous activity from a reliable
counterpart source at other wavelengths.

The 
final verification is basically divided into (1) a
visual inspection of the \textit{AGILE} sky maps (see Section
\ref{man1}), and (2) 
a data analysis by human intervention (see Section
\ref{man2}).

\subsection{Quick Look of Sky Maps}
\label{man1} When an automatic candidate alert is received, the first step 
performed during the monitoring activity is the \abp{quick look} of the
data products (e.g., maps and light curves)  through web pages
accessible via \abp{a} web browser \abp{for} personal computers
or smartphones and via the \textit{AGILEScience App}.

\subsubsection{The  AGILEScience App}
\label{SAPP}

To support the management of candidate alerts, an iOS App
(the \textit{AGILEScience App}) has been developed at IASFBO by
the AGILE Team in collaboration with the University of Modena and
Reggio Emilia. This app   consists of (1) a public section
with outreach purposes,  (2) a private section, and (3) a
notification system that sends the \textit{AGILE}-GRID alerts to the AGILE Team. Furthermore, the  App  notification system
generates also alerts (to everyone, not only to AGILE Team) when a
new Astronomical
Telegram\footnote{http://www.astronomerstelegram.org} or GCN
circular\footnote{http://gcn.gsfc.nasa.gov} is published.

In the \textit{public section},  the last available full sky \gray
map can be seen by the public for scientific and outreach
purposes \abp{in order} to follow the evolution of the \gray sky as 
detected by \textit{AGILE} (the map is updated every \textit{AGILE} orbit). This means that also the
non-professional astronomer can follow the  \gray sky
 'in real-time'. All  the public content of the \textit{AGILEScience
App} is also available through a web browser  \abp{for} Android mobile
devices\footnote{http://agile.iasfbo.inaf.it}.

The \textit{private section}   is  password protected and shows
the  analyzed data, in particular, (1) zoomable   full sky maps
and the analyzed regions overlapped with the automated results
provided by \sasnospace, and (2) data quality reports provided by \qlsnospace.

The App for mobile devices is deeply integrated into a
scientific ground segment and it is used for daily scientific
activities. When an automatic candidate alert is received
through the notification system,  the private section of the App
can be used to access the \textit{AGILE}-GRID maps to check the  automated
results or the quality of the data.

\subsection{Final Analysis} 
\label{man2}

To confirm or to improve the automated analysis results before the publication of an ATel, \abp{an
analysis by human intervention 
is performed} {\mtt for a subclass of candidate transient  events 
following the same data analysis strategy as reported in
Section \ref{sec31}. This
analysis is
performed taking into account all known sources in the transient
region,
\abp{and uses} the most updated version of the \textit{AGILE} \gray source
catalog.
For the known sources, we use fixed positions and fluxes in
the likelihood analysis.
 We then add to  this ensemble of 
 known sources a \abp{point-like} source representing the transient
 candidate initially using the  position
as determined by the \sasnospace. We then perform the likelihood
analysis, \abp{initially setting} the position and flux of this transient candidate \abp{free}.

In addition,  we 
first estimate the $g_{gal}$ and $g_{iso}$ parameters with
a longer timescale integration (typically 15 days of integration).
We
then  fix them for
the short timescale analysis, assuming that these 
parameters do not vary significantly on
timescales \abp{on orders} of hours-days.
Typically, the integration time we use for this
final analysis is the same as for the automatic analysis
(\abp{one to two} days). However, we vary the integration time for timescales
smaller than \abp{one} day to find the \gray flare peak.

\section{Statistical Significance of IASFBO SAS Pipeline Data Analysis Technique}
\label{S5}

\abp{In this section, we discuss the methods used} to determine the
statistical significance of our procedure. Our method  is
characterized by adding data for the map analysis by a ``1-orbit
sliding window offset". This procedure implies that the 15 maps
generated every day are not independent from one another, and our
statistical analysis has to take this fact into account.
This is a very different approach
than \abp{the one taken by} \cite{Bulgarelli_2012}, \abp{which}  required a new evaluation of
the \abp{pretrial} significance of a detection and new simulations. In
the end, final results are quite similar but the hypothesis of
\cite{Bulgarelli_2012} and of this \abp{article} are  different from a
statistical point of view. For the same reason,  the evaluation of
the \abp{posttrial} probability of \cite{Bulgarelli_2012} has  to be
\abp{redone} and is not valid in this context.

On the basis of the current \textit{AGILE}-GRID performance in
orbit \cite{Chen_2013}, \abp{which is}  characterized by a relatively large
effective area at 100 MeV and a very large FoV,
 we determine the range of the \abp{pretrial} test statistic $T_s$ in
terms of the probability of false detections (the $p$-value) of the
automated analysis procedures. The $T_s$ is defined as -2~ln ${L_0}/{L_1}$ where $L_0$ and $L_1$ are the maximum value of the likelihood function for the null hypothesis and the alternative hypothesis, respectively.

The data is analyzed  each orbit, \abp{and we reanalyze} the data of the 48-hr period (in ``spinning" mode)
ending with the current orbit. This procedure
implies a sliding window offset by one orbit. The specific search
(whether a ``blind" search or a search for a specific source) has
a direct influence on the proper statistical treatment.

\subsection{Statistical Significance of the Blind Search Procedure for Unknown Transient Sources}

First we address the \textit{blind search procedure}: the flaring source
is unknown, and the position and flux parameters of each candidate
are allowed to be free and optimized with respect to the input
data. The  starting $(l, b)$ position is determined with
the already described method called \textit{spotfinder}.

The determination of the likelihood ratio distribution in the case
of the null hypothesis is used to evaluate the occurrence of false detections.
To evaluate the $p$-value of the sliding window approach, we
performed simulations of  empty Galactic regions (i.e., without
steady or flaring source). The simulated observations were generated by adding
Poisson-distributed 
counts in each pixel \abp{while} considering the exposure level, the
Galactic diffuse emission model, and the isotropic diffuse
intensity. To simulate the sliding window, we simulated 60
exposure and counts maps of $96^\prime$ of \textit{AGILE} orbit for each run
(corresponding to four days of observation, 15 orbits for each
day). We then added the first 30 of these maps to obtain a
two-day observation map; \abp{we then} removed the first and added the
next $96^\prime$ map, and so on. Each \abp{two-day} generated maps (counts,
exposure, and Galactic emission maps) have been analyzed using the
same procedure as the
on-orbit data. Simulated data 
are
 generated using the \textit{AGILE}-GRID instrument response
functions I0023 \cite{Chen_2013}. The parameters used in the
simulation are $g_{gal}$ = 0.6 and $g_{iso}$ = 8 (typical
of the GRID data processing). During the analysis the spectra of
all sources in the field are kept fixed.

We 
then calculate the $p$-value distribution, analyzing at the
same time \abp{one, three and eight} candidate flares (as upper limits) in
an empty field, with the flux and position of each source allowed
to vary. Table \ref{table_1} reports the performed simulation and
related parameters.

The probability that the result of a trial in an empty field has
$T_s \geq h$ is
\begin{equation}
P(T_s \geq h) = \int_{h}^{+\infty} \varphi (x) dx
\label{eq_A}
\end{equation}
which is also called the $p$-value $p=P(T_s \geq h)$. This is the
 pre-trial type-1 error (a false positive, rejecting the
null hypothesis when in fact it is true); {\mmt the} ``$p$-value"
assigned to a given value of a random variable is defined as the
probability of obtaining that value or larger when the null
hypothesis is true. The resulting $p$-value distributions are shown
in Figures \ref{fig_ts1}$-$\ref{fig_ts3}.
We fit  these distribution with the function,
\begin{equation}
\kappa''(T_s) = \left\{
\begin{array}{ll}
\delta & \mbox{if }   T_s<1  \\
\eta_1 \chi^2_{N_1}(T_s)  & \mbox{ if $T_s\ge 1$ and $T_s\le t_{lcl}$ } \\
\eta_2 \chi^2_{N_2} (T_s - t_{lcl}) & \mbox{ otherwise }
\end{array}
\right.
\label{eqn_posfree}
\end{equation}
\noindent where $t_{lcl}$ is a threshold that \abp{enables} the location
of the contour level: a typical value is $t_{lcl} = 5.99147$, \abp{which} corresponds to a 95\% confidence level for two degrees of freedom.
We find
that $N_1$ = 1 (if $T_s < t_{lcl}$, no optimization of the
position takes place, and the only free parameter is the 
source flux), $N_2 = 5$, and the remaining parameters depend on
the hypothesis; Table \ref{table_ef2} reports results for some
cases. As already stated, similar results have been obtained 
by  the analysis of an empty Galactic region with
no
sliding window 
\citep{Bulgarelli_2012} but under different hypothesis.

Fitting the distribution of the $g_{gal}$ and $g_{iso}$
parameters with a Gaussian distribution,  we obtain $g_{gal} =
0.68 \pm 0.15$ and $g_{iso} = 8.1 \pm 2.7$. Similar results have
been obtained for \abp{eight} sources in the ensemble of models.

\subsection{Statistical Significance of the Search Procedure for Transient Emission from Known Sources}

In the following \abp{section,} we address \textit{the search for a specific
transient source} using a list of known \gray sources in the
region. The main difference between this procedure and the blind
search procedure already described is that the source
positions  are kept fixed, and only the fluxes are allowed
to vary.
Fitting these distribution with the function,
\begin{equation}
\kappa'(T_s) = \left\{
\begin{array}{ll}
\delta   &\mbox{ if $T_s<1$}, \\
\eta \chi^2_{N_1} (T_s) &\mbox{ otherwise, } \\
\end{array}
\right.
\label{eqn_fit1}
\end{equation}
\noindent we {\mtt find} that $\delta = 0.86 \pm 3.4\times
10^{-4}$, $N_1 = 1$, and $\eta=0.45 \pm  4.5\times 10^{-4}$.

Table \ref{table_final} reports the correspondence between
$p$-values and $T_s$ values for different numbers of point sources
in the ensemble of models, and for two search procedures used in
the \sasnospace. Figure \ref{fig_ts1} shows the $p$-value distribution for
one candidate in the ensemble of models with the  position
kept fixed (brown line).

\section{The Most Relevant Results}
\label{S6}

The \textit{AGILE}  Alert System described in this
paper 
led to issuing more than 
90 Astronomical
Telegrams (ATels) in about seven years of
operations. 
Among the most noticeable alerts \abp{that warrant mentioning are as follows:}
\begin{enumerate}

 \item The first detection of   transient \gray emission from Cygnus X-3 in the
energy range of 100 MeV$-$50 GeV \citep{Tavani_2008}, \abp{which was} confirmed by the \textit{Fermi}/LAT collaboration in
\cite{abdo2b}
and reported in \cite{Tavani_2009b} and
\cite{Bulgarelli_2012_a, Piano_2012} .

\item The discovery of
\gray 
flares from  the Crab Nebula in 2010 \abp{September}
\citep{Tavani_2010} (confirmed by \textit{Fermi}-LAT within 1 day, see
\cite{Buehler_2010}). 
See Section
\ref{crab} for details. The first detection of a Crab Nebula  flare was made in 2007 \abp{September} by the \sasnospace.
\item The first ATel that alerted the astrophysical
community of the extraordinary activity of the blazar 3C454.3 \abp{in
2010 December}, \abp{which was} in addition to the detection, early in the
mission (2007) and at a later stage (2009 and 2010), of very bright
\gray emission \cite{Vercellone2010ApJ_P3,Vercellone_2011};
see Section \ref{3c454}.

\item The detection of many \gray flares from blazars.

\end{enumerate}

In most cases,  the evolution of the \gray flare 
could be followed in real-time  (with a delay of 
approximately  \abp{two} hours, because of the
 orbit-by-orbit  integration provided by the \sas system).
The  procedure and fast link (through cellular telephones) to
the \gray maps and processing results
turned out to be crucial in a number of occasions.
Another   key factor  is the
 \abp{team} organization in the management of the alerts.

\abp{Tables \ref{tableA1} and \ref{tableA2}} list  the
\gray transient sources published in Astronomical Telegrams.
Figure \ref{fig_atel1} and \ref{fig_atel2} show the position and
classification of the published \textit{AGILE}-GRID ATels in ``pointing" and
``spinning" mode overlapped to the related exposure maps.

\subsection{The 2010 Crab Flare Case}
\label{crab}

The Crab source (pulsar $+$ Nebula) is usually  characterized
by 
a mean flux of $F = (2.2 \pm 0.1) \times 10^{-6}$ $\phcmsec$ for
\enmev at a significance of $\sqrt{T_s}$ = 30.0. This
result is obtained  with data from \abp{2007 July to 2009 October}, taking
into account the diffuse \gray background with Galactic and
isotropic components and  considering all nearby sources with a
fixed flux.
Figure \ref{fig_crab2} 
shows "the sliding lightcurve" of the 2010 September Crab
Nebula flare as seen by the \sas described in this paper.

The first alert from a source positionally consistent with
the Crab  was generated by the pipeline for a source
intensity exceeding by 1 $\sigma$ the mean flux level (see
the time segment called "1.a" in Figure \ref{fig_crab2}).
An automated message was sent
by  e-mail and SMS
on \abp{2010} September 20, 02:04:04 UT (the yellow arrow 1.b in
figure reports the alert generation). The Crab Nebula  
reached its maximum flux in \textit{AGILE} data (see  the time
segment ``2.a" in Figure \ref{fig_crab2}) during the
integration time from 2010 September 19 01:54:43 UT 
until 2010 September 20 23:47:51 UT. The 
corresponding
alert was 
sent by e-mail and SMS
on 2010 September 21 02:00:54 UT (see red arrow 2.b), about
two hours after the maximum of the physical phenomena. 
During this period, the standard \textit{AGILE}-GRID processing \abp{showed}
\gray flux levels from the Crab region of total significance
above $8 \, \sigma$, a highly unusual situation for \abp{two}-day
integrations in spinning mode. The Astronomical Telegram 2855
announcing the existence of a flaring source positionally
consistent with the Crab Nebula  was posted \abp{on} 2010 September 22
\abp{at} 14:45:00 UT \cite{Tavani_2010}.
\textit{AGILE} and \textit{Fermi}-LAT produced five ATels within the short time of a few days.
This procedure and alert system  resulted 
in \abp{a} very  fast communication to the astrophysical community
of the  discovery of  \gray variability from the
Crab Nebula.

\subsection{The Brightest Gamma-Ray Blazar in the Sky, 3C~454.3, \textit{the Crazy Diamond}}
\label{3c454}

The flat spectrum radio quasar 3C~454.3 (PKS 2251$+$158; $z$ = 0.859) is the brightest
\gray (0.1--10 GeV) blazar detected after the \abp{launches} of the \textit{AGILE}
and {\it Fermi}    satellites.
Since 2007, \textit{AGILE} detected and investigated several \gray flares, as reported in
\citep{Vercellone2010ApJ_P3}.

The extremely rapid analysis of the \textit{AGILE} \gray data allowed us to trigger several
Target of Opportunity (ToO) observations with both ground- and space-based Observatories,
such as the  GLAST-AGILE Support Program (GASP; \cite{vil08,vil09})  of the Whole
Earth Blazar Telescope\footnote{http://www.oato.inaf.it/blazars/webt/} (WEBT;
radio, optical and infrared band); the {\it Swift} Satellite \citep[][optical, ultra-violet, soft and hard X-ray]{Gehrels2004:swift};
INTEGRAL \citep[][soft and hard X-ray]{Winkler2003AA};
and MAGIC \citep[][above 100~GeV]{Aleksic2012APh}.
It is of paramount importance to alert the community during the onset of a \gray flare
(whose duration can vary from two up to several days), in order to be able to catch the peak
of the \gray emission in a \abp{multifrequency} fashion.

Such an unprecedented, panchromatic, and almost simultaneous coverage of this bright source
allowed us to  establish a possible correlation between the \gray
(0.1 --10 GeV) and the optical ($R-$band) flux variations with no time delay, or with a
lag of the former with respect to the latter of about half a day.
Moreover, the detailed physical modeling of the \abp{spectral energy distributions}
when 3C~454.3 was at different flux levels provided an interpretation of the emission mechanism responsible
for the radiation emitted in the \gray energy band, \abp{which has} assumed to be inverse Compton scattering of photons from
the \abp{broad-line} region (BLR) clouds off the relativistic electrons in the jet, with a bulk Lorentz
factor of $\Gamma \sim 20$.

On 2010 November 20, 3C~454.3 reached a peak flux ($E>$100\,MeV) of
$F_{\gamma}^{\rm p} = (6.8\pm1.0)\times 10^{-5}$\,ph\,cm$^{-2}$s$^{-1}$\, on a time scale
of about 12 hr, more than a factor of 6 \abp{times} higher than the flux of the brightest
steady \gray source, the Vela pulsar \citep{Vercellone_2011}.
The \textit{AGILE} rapid alert system allowed us to follow this event and trigger several ToOs for
about one month. Such a dense \abp{multifrequency} coverage allowed us to detected not
only the major \gray flare, but also a peculiar  \gray {\it orphan} optical
flare about 10 days prior to the major \gray flare. This puzzling behavior challenges
the idea of a uniform external photon field and is still under investigation,  \abp{see V. Vittorini et al. (2014, in preparation)}.

\section{Conclusions}
\label{sec:conclusion}
In this paper we described the main features of the fast \gray data processing of the \textit{AGILE} mission. In particular, we
focused on the 'spotfinder' algorithm, the optimization of software tools, the data link from the satellite to data processing centers, the orbit-by-orbit data analysis, and the statistical characterization of the data analysis system.
An important part of the data processing is  the extensive use of mobile technologies
coupled with the simultaneous
implementation of two independent
pipelines of the  AGILE Alert System.

Identifying unexpected transient astrophysical
events within a very short time is of crucial
importance for high-energy astrophysics.  The \textit{AGILE}
Alert System has demonstrated to be quite successful in source
detection and rapid alert capability. The AGILE mission and the
scientific community have certainly  benefited from its
implementation, which maximizes the scientific return of \gray
observations.

Our work can be important for future high-energy
astrophysics instruments operating at \gray energies. Very
efficient data transmission to the processing center and an
orbit-by-orbit data analysis are the crucial ingredients for an
efficient scientific Ground Segment. The AGILE-GRID Alert System
demonstrates that with  a
proper choice of resources, the task of automatic
alerting for transient cosmic sources within \abp{two to three} hours is
possible. We expect this task to become an  important requirement
for future high-energy astrophysics instruments both in space and
on the ground.

\acknowledgments The \textit{AGILE} Mission is funded by the Italian Space
Agency (ASI) with scientific and programmatic participation by the
Italian Institute of Astrophysics (INAF) and the Italian Institute
of Nuclear Physics (INFN).
Our research is partially
supported by the ASI grants I/042/10/0 and I/028/12/0.




\clearpage

\begin{figure}[!htb]
\centering
\includegraphics[width=16 cm]{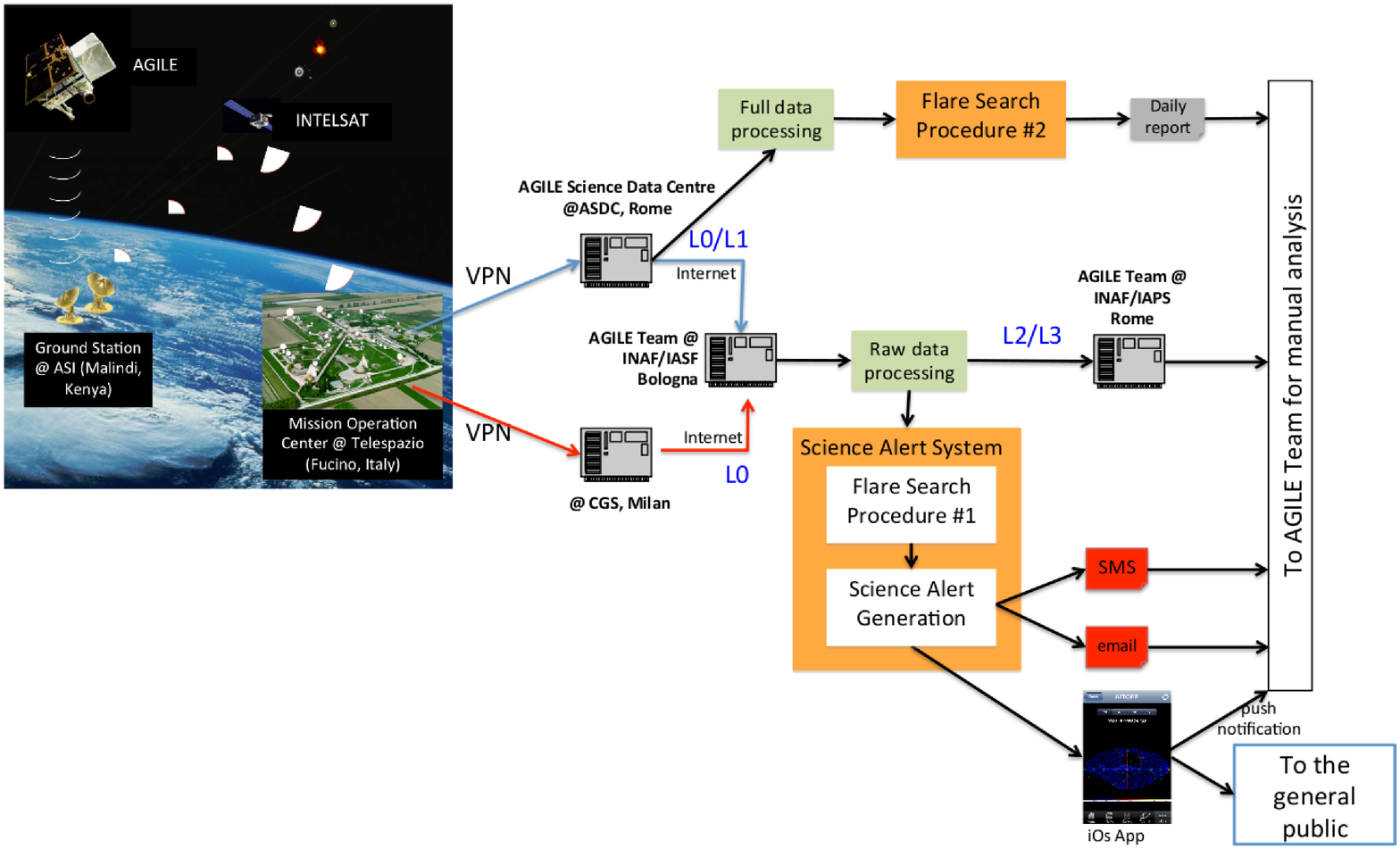}
\centering \caption {  General architecture of the Ground Segment with details of the online AGILE Alert System (green boxes indicate the data processing, and orange boxes indicate the most specific pipelines for $\gamma$-ray flare search and alert generation) in terms of nodes of elaboration and functional blocks. The arrows indicate the data flow between different nodes and functional blocks. 
 }
\label{fig_F1}
\end{figure}

\clearpage

\begin{figure}[!htb]
\centering
\includegraphics[width=6 cm]{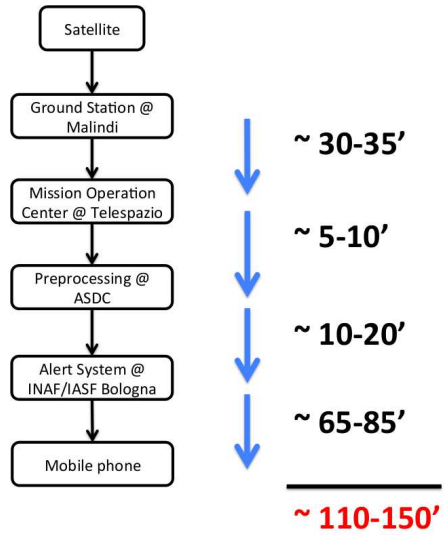}
\centering \caption { Data transfer time of the Ground Segment from the \textit{AGILE} satellite to SMS of scientific alerts to mobile phones.  }
\label{fig_time}
\end{figure}

\clearpage

\begin{figure*}[!htb]
\centering
\includegraphics[width=16 cm]{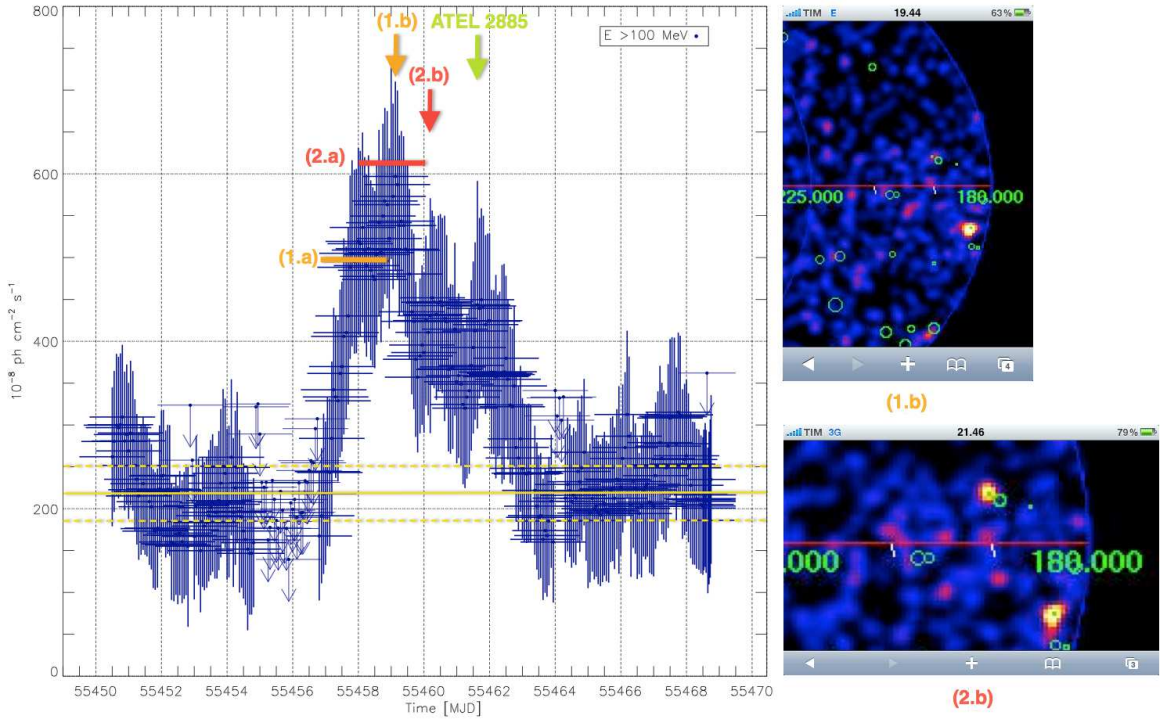}
\centering \caption {$96^{\prime}$ sliding light curve (with two-day integration time)
of the 2010 September Crab nebula flare as seen by \sasnospace. Errors are  $1\sigma$, and
time is given in MJD. The yellow lines show the average Crab flux and the
$3\sigma$ uncertainty range. 1.a and 1.b (in orange) are, respectively, the detected flux and the time of the alert generation by the \sas when Crab nebula reaches a flux level that exceeds $1\sigma$ the mean flux level; on the right are the counts map of 1.b as seen by \sasnospace. 2.a and 2.b (in red) are related to the maximum flux level reached; on the right are the counts map of 2.b as seen by \sasnospace. The green arrow indicates the time that the Astronomical Telegram was posted. }
\label{fig_crab2}
\end{figure*}

\clearpage

\begin{figure}[!htb]
\centering
\includegraphics[width=16 cm]{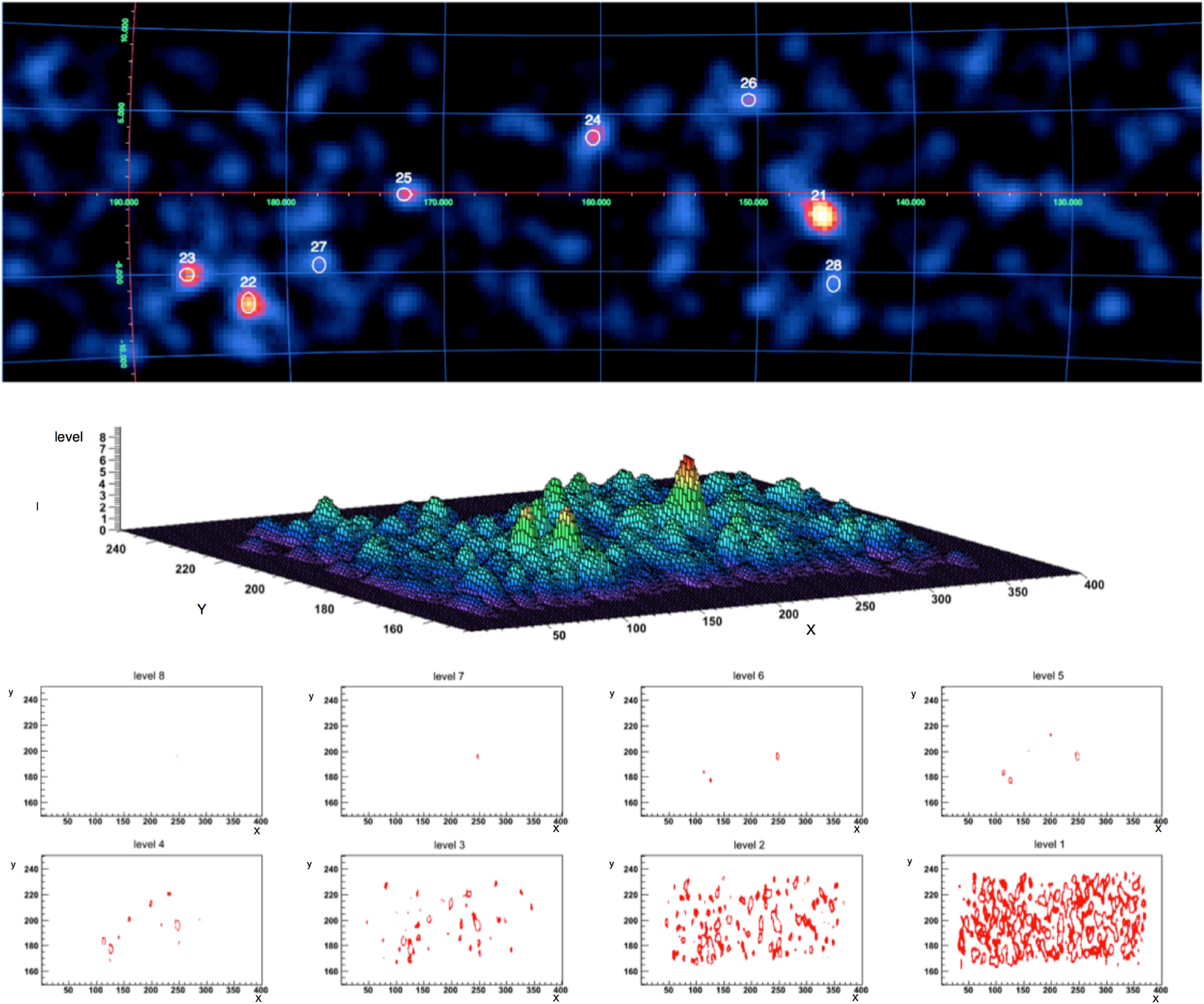}
\centering \caption { \textit{spotfinder} algorithm for the extraction of connected regions into a multilevel \gray map. Top panel: a region of the Galactic plane used as an example; also shown in the same picture are the  connected component regions identified by \textit{spotfinder}. Second panel: the same sky region normalized in eight levels. Third panel: the eight images used to grow the connected regions. }
\label{fig_spotfinder}
\end{figure}

\clearpage

\begin{table}
\begin{center}
\caption{Performed Simulations for the Evaluation of Statistical Significance of the Blind Search Procedure for Unknown Transient Sources. }
\label{table_1}
\begin{tabular}{ccc}
\tableline\tableline
Source Position  & Candidate Flares & Simulated Maps $\times 10^6$  \\
\tableline
 Free & 1  &  10  \\
 Free &  3 &   5 \\
 Free &  8 &   5 \\
 Fixed &  1 &   8  \\
\tableline
\end{tabular}
\end{center}
\end{table}

\clearpage

\begin{figure}[!htb]
\centering
\includegraphics[width=14 cm]{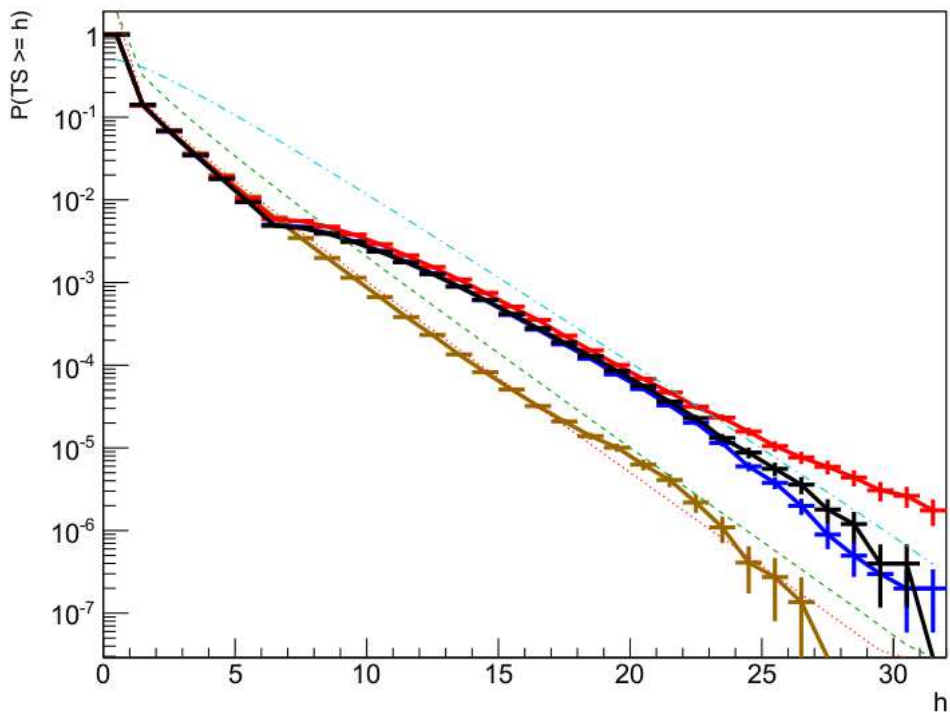}
\centering \caption { Histograms are the \textit{p}-value distributions for an empty Galactic region with a different number of candidate flare in the ensemble of models. Each histogram shows the \textit{p}-value distribution of the first candidate in the ensemble of models (the blue histogram has one candidate, the red histogram has three candidates, and the black histograms has eight candidates). The flux and position of the sources are left free; $g_{\rm gal}$ and $g_{\rm iso}$ parameters are free. The brown histogram shows the \textit{p}-value distribution for one candidate in the ensemble of models with its position kept fixed. The red dotted line is the $\frac{1}{2} \chi^2_1$ theoretical distribution, the green dashed line is the $\chi^2_1$ theoretical distribution, and the cyan dot-dashed line is the   $\frac{1}{2} \chi^2_3$ distribution. }
\label{fig_ts1}
\end{figure}

\clearpage

\begin{figure}[!htb]
\centering
\includegraphics[width=14 cm]{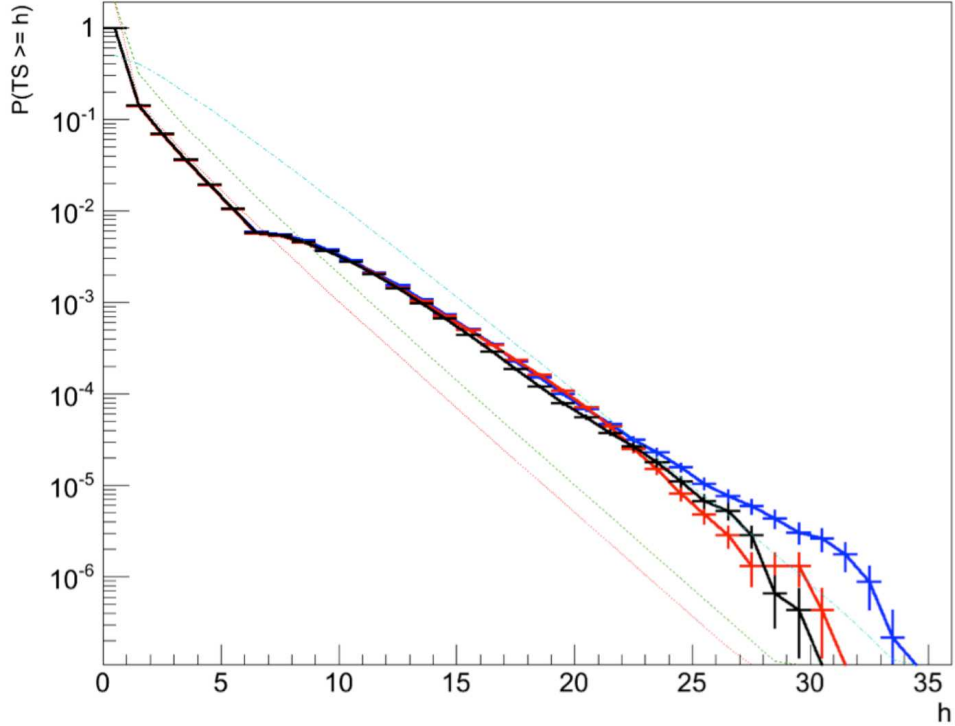}
\centering \caption { Histograms are the \textit{p}-value distributions for an empty Galactic region when there are three sources in the ensemble of models. The blue histogram is the distribution of the first source, the red histogram is the distribution of the second source, and the black histogram is the distribution of the third source. The flux and position of the sources  are left free; $g_{\rm gal}$ and $g_{\rm iso}$ parameters are free. The red dotted line is the $\frac{1}{2} \chi^2_1$ theoretical distribution, the green dashed line is the $\chi^2_1$ theoretical distribution, and the cyan dot-dashed line is the   $\frac{1}{2} \chi^2_3$ distribution.  }
\label{fig_ts2}
\end{figure}

\clearpage

\begin{figure}[!htb]
\centering
\includegraphics[width=14 cm]{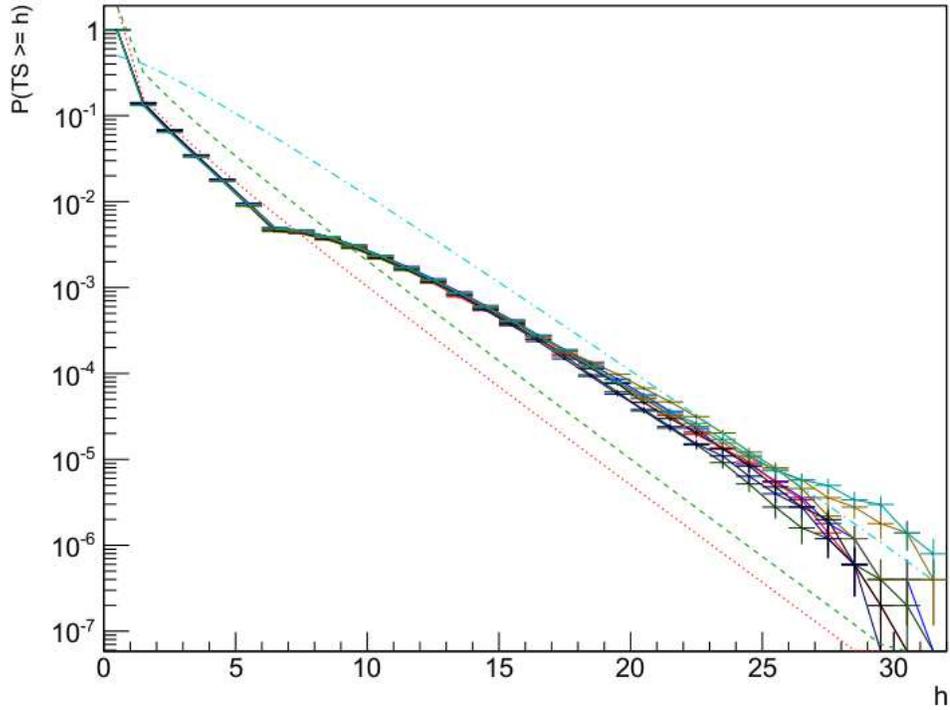}
\centering \caption { Histograms are the \textit{p}-value distributions for an empty Galactic region when there are eight sources in the ensemble of models. The flux and position of the sources are left free; $g_{\rm gal}$ and $g_{\rm iso}$ parameters are free. The red dotted line is the $\frac{1}{2} \chi^2_1$ theoretical distribution, the green dashed line is the $\chi^2_1$ theoretical distribution, the cyan dot-dashed line is the $\frac{1}{2} \chi^2_3$ distribution.}
\label{fig_ts3}
\end{figure}

\clearpage

\begin{table}
\begin{center}
\caption{Fitting Results for the Parameters Reported in Equation (\ref{eqn_posfree}) for a Different Hypothesis Formulation.}
\label{table_ef2}
\begin{tabular}{lcccc}
\tableline\tableline
Ensemble of models & \textbf{$\delta \times 10^{-4}$} & $\eta_1 \times 10^{-4}$ & $\eta_2$  \\
\tableline
One candidate flare & $0.86 \pm 2.9$ & $0.45 \pm 3.9$ & $4.8 \times 10^{-3} \pm 2.1\times 10^{-5}$ \\
Three candidate flares - first source & $0.86 \pm 4.3$ & $0.45 \pm 5.8$ & $5.7 \times 10^{-3} \pm 3.5\times 10^{-5}$ \\
Three candidate flares - third source & $0.85 \pm 4.3$ & $0.46 \pm 5.8$ & $5.5 \times 10^{-3} \pm 3.4\times 10^{-5}$ \\
Eight candidate flares - first source & $0.85 \pm 4.1$ & $0.46 \pm 5.6$ & $4.7 \times 10^{-3} \pm 3.0\times 10^{-5}$ \\
Eight candidate flares - last source & $0.87 \pm 4.1$ & $0.42 \pm 5.8$ & $4.7 \times 10^{-3} \pm 3.0\times 10^{-5}$ \\
\tableline
\end{tabular}
\end{center}
\end{table}

\clearpage

\begin{table}
\begin{center}
\caption{Correspondence between \textit{p}-value and $T_s$ Value for a Different Number of Point Sources in the Ensemble of Models and for Two Search Procedures Used in the \sasnospace.}
\label{table_final}
\begin{tabular}{|l|l|l|lll|}
\tableline\tableline
\textbf{$\sigma$ pre-trial} & \textbf{p-value} & \textbf{Fixed position} & & \textbf{Free position} &  \\
 & & 1 candidate & 1 candidate  & 3 candidates &    8 candidates \\
\tableline
4 & $3.17 \times 10^{-5}$ & 15.8 & 22.2 &   22.4-22.5   & 22.1-22.8\\
5 & $2.86 \times 10^{-7}$ & 24.8 & 33.0 & 33.2-33.3  & 32.8-33.5\\
6 & $9.21 \times 10^{-10}$ & 35.8 & 45.4    & 45.6-45.7 &   45.2-45.9\\
\tableline
\end{tabular}
\end{center}
\end{table}

\clearpage

\begin{deluxetable}{cccccc}
\tabletypesize{\scriptsize}
\tablewidth{0pt}
\tablecaption{\textit{AGILE}-GRID ATels in ``pointing" mode.}
\tablehead{
\colhead{Positionally Consistent with}      &
\colhead{Object Class}          & \colhead{ATels} &  \colhead{Start (UT)} & \colhead{Stop (UT)} & \colhead{ATel ID}}
\startdata
\multirow{8}{*}{3C 454.3} &  \multirow{8}{*}{Blazar}& \multirow{8}{*}{8}  & 2007 Jul 24 14:30& 2007 Jul ­27 05:27& 1160\\
 &  & &  2007 Jul ­24 14:30& 2007 Jul ­30 11:40& 1167\\
 &  & &  2007 Nov 02 13:50& 2007 Nov 12 17:01& 1278\\
 &  & &  2007 Nov ­02 12:00& 2007 Nov ­22 18:33  & 1300\\
 &  & &  2008 May ­24 03:10 & 2008 May ­27 04:17 & 1545\\
 &  & & 2008 Jun ­15 10:46& 2008 Jun 16 07:11& 1581\\
 &  & &  2008 Jun ­25 03:00 & 2008 Jun ­26 03:00 & 1592\\
 &  & &  2008 Jul ­25 18:00 & 2008 Jul ­28 03:00 & 1634\\
\hline
\multirow{4}{*}{PKS 1510-089}  &  \multirow{4}{*}{Blazar} &  \multirow{4}{*}{4}  & 2008 Mar ­18 03:00 & 2008 Mar ­20 03:00 & 1436\\
& & & 2009 Mar ­08 14:00 & 2009 Mar 10 4:00 & 1957\\
& & & 2009 Mar 12 07:00 & 2009 Mar 13 05:00 & 1968\\
& & & 2009 Mar ­18 05:45& 2009 Mar 19 05:33 & 1976\\
\hline
PKS 1830-211  &  Blazar & 1  & 2009 Oct 12 00:03& 2009 Oct ­13 04:57& 2242\\
\hline
Mkn 421  &  Blazar &  1  &2008 Jun ­09 17:02 & 2008 Jun ­15 02:17& 1583\\
\hline
3EG J0721+7120  &  Blazar &  1  & 2007 Sep ­10 13:50& 2007 Sep ­20 10:13&1221\\
\hline
3EG J1410-6147 &  SNR &  1  & 2008 Feb ­21 06:00& 2008 Feb ­22 07:30&1394\\
\hline
W Comae  &  Blazar & 1  & 2008 Jun 09 17:02& 2008 Jun ­15 02:17& 1582\\
\hline
\multirow{3}{*}{AGLJ2021+4029} & \multirow{3}{*}{Unidentified}  & \multirow{3}{*}{3} & 2008 Apr ­27 01:39& 2008 Apr ­28 01:27& 1492\\
& & & 2008 May ­22 06:00 & 2008 May ­27 06:00& 1547\\
& & & 2007 Nov ­02 12:00 & 2008 May ­01 00:00& 1585\\
 \hline
AGL J2021+4032 & Unidentified &  1  & 2008 Nov 16 14:33& 2008 Nov ­17 14:22& 1848\\
\hline
AGL J2030+4043 & Unidentified  &  1  & 2008 Nov 02 20:43 & 2008 Nov ­03 20:32&1827\\
\hline
AGL J0229+2054 & Unidentified  &  1  &2008 Jul 30 15:34 & 2008 Jul ­31 15:23& 1641\\
 \hline
AGL J1734-­3310  & Unidentified &  1  & 2009 Apr 14 00:00& 2009 Apr ­15 00:00& 2017\\
\hline
\enddata
\label{tableA1}
\end{deluxetable}

\clearpage

\begin{deluxetable}{cccccc}
\tabletypesize{\scriptsize}
\tablewidth{0pt}
\tablecaption{\textit{AGILE}-GRID ATels in ``spinning" mode.}
\tablehead{
\colhead{Positionally Consistent with}      &
\colhead{Object Class}          & \colhead{ATels}  &  \colhead{Start (UT)} & \colhead{Stop (UT)} & \colhead{ATel ID} }
\startdata
\multirow{6}{*}{3C 454.3}& \multirow{6}{*}{Blazar} & \multirow{6}{*}{6} & 2009 Nov ­26 01:00 & 2009 Dec ­2 08:30 & 2322\\
 &   & & 2009 Dec ­2 06:30 & 2009 Dec ­3 08:30 & 2326\\
   &  & & 2010 Oct ­28 06:00 & 2010 Oct 31 06:00 & 2995 \\
 &   & &2010 Nov ­16 06:50 & 2010 Nov ­17 09:15 & 3034\\
 &   & & 2010 Nov ­18 03:15 & 2010 Nov ­19 09:15 & 3043\\
 &  & & 2010 Nov 21 04:50 & 2010 Nov 22 04:50 & 3049\\
\hline
\multirow{5}{*}{PKS 1510-089} &  \multirow{5}{*}{Blazar} & \multirow{5}{*}{5}  &
2010 Jan 11 10:30& 2010 Jan ­13 10:30 & 2385\\
& & &2011 Jul 2 04:30& 2011 Jul ­4 04:30& 3470\\
& & & 2012 Jan ­28 12:00& 2012 Feb 1 12:00 &3907\\
& & &2012 Feb ­14 04:00 & 2012 Feb 17 04:30& 3934\\
& & & 2013 Sep 22 12:00  & 2013 Sep 24 12:00 & 5422\\
\hline
\multirow{2}{*}{PKS 1830-211}  &  \multirow{2}{*}{Blazar} & \multirow{2}{*}{2}  & 2009 Nov ­20 17:00& 2009 Nov 22 17:00 &2310\\
& & & 2010 Oct 15 00:00& 2010 Oct ­17 00:00& 2950\\
\hline
\multirow{2}{*}{PKS 0402-362}  &  \multirow{2}{*}{Blazar}  &  \multirow{2}{*}{2}  & 2010 Mar ­13 12:00 & 2010 Mar 16 15:00 & 2484\\
 & & & 2011 Aug 8 04:10 & 2011 Aug ­10 09:04& 3544\\
\hline \multirow{3}{*}{PKS 1222+216}  &  \multirow{3}{*}{Blazar}  &  \multirow{3}{*}{3}  &2009 Dec ­13 06:00 & 2009 Dec ­15 06:00 &2348\\
 & & & 2010 May ­23 14:30 & 2010 May ­26 05:22 & 2641\\
 & & & 2010 Jun ­17 09:20 & 2010 Jun ­19 09:30 & 2686\\
\hline
BL Lac  &  Blazar  & 1 & 2011 May 27 11:57 & 2011 May ­29 11:30 & 3387\\
\hline
S41749+70  &  Blazar  & 1 & 2011 Feb ­26 00:00& 2011 Mar ­1 04:00 & 3199\\
\hline
  \multirow{2}{*}{4C +3841}  &  \multirow{2}{*}{Blazar}  &  \multirow{2}{*}{2} & 2012 Sep ­15 22:00& 2012 Sep ­18 10:30 & 4389\\
& & & 2013 Jul ­25 06:00& 2013 Jul ­27 15:00& 5234\\
\hline
  2FGL J1823.8+4312  &  Blazar  &  1  & 2012 Jun 3 04:00& 2012 Jun ­5 04:00& 4153\\
\hline
  2FGL J1127.6+3622  &  Blazar  &  1  & 2011 Nov ­17 00:00& 2010 Nov ­19 00:00 &3858\\
\hline
  PKS J2329-4955  &  Blazar  &  1  & 2010 Nov 1 00:00& 2010 Nov ­5 03:00&3008\\
\hline
  PKS 1830-211  &  Blazar  &  1  & 2010 Oct ­15 00:00& 2010 Oct ­17 00:00 & 2950\\
\hline
  PKS 2142-758  &  Blazar  &  1  & 2010 Apr ­10 11:30 & 2010 Apr ­12 21:00& 2551\\
\hline
  PKS 0537-441  &  Blazar  &  1  & 2010 Feb 18 00:00& 2010 Feb ­23 10:30& 2454\\
\hline
  3C 273  &  Blazar  &  1  & 2010 Jan ­6 16:50& 2010 Jan ­8 07:50& 2376\\
\hline
  PKS 2233-148  &  Blazar  &  1  & 2012 Jun 3 10:00& 2012 Jun ­5 22:00& 4154\\
\hline
  PMN J0948+0022  &  Seyfert 1  &  1  & 2011 Jun ­20 07:30& 2011 Jun ­22 11:20& 3448\\
\hline
 \multirow{2}{*}{Cygnus X-1}  &  \multirow{2}{*}{Binary}  &  \multirow{2}{*}{2}  & 2010 Mar ­24 02:24& 2010 Mar ­25 01:01& 2512\\
 & & &2010 Jun 30 10:00& 2010 Jul ­2 10:00&2715\\
\hline
  \multirow{6}{*}{Cygnus X-3}  &  \multirow{6}{*}{Binary}  &  \multirow{6}{*}{6}  & 2010 May ­7 14:53& 2010 May ­9 17:19& 2609\\
  & & & 2010 May ­25 19:10& 2010 May ­27 17:04& 2645\\
  & & & 2011 Jan ­27 20:00& 2011 Feb 1 11:00& 3141\\
  & & & 2011 Feb ­6 05:00& 2011 Feb ­8 09:00& 3151\\
  & & & 2011 Mar 20 00:00& 2011 Mar 20 00:00& 3239\\
  & & & 2011 May 28 07:58& 2011 May 29 06:02& 3386\\
\hline
  PSR B1259-63 &  Pulsar  &  1  & 2010 Aug ­2 06:400& 2010 Aug ­4 04:00& 2772\\
\hline
  \multirow{6}{*}{Crab Nebula}  &  \multirow{6}{*}{SNR}  &  \multirow{6}{*}{6}  & 2010 Sep ­19 00:10& 2010 Sep ­21 00:10&2855\\
  & & & 2011 Apr ­9 23:45& 2011 Apr ­13 23:45& 3282\\
  & & & 2011 Apr 15 10:40 & 2011 Apr 16 10:38& 3286\\
  & & & 2013 Mar 3 05:30& 2013 Mar ­4 11:00& 4856\\
  & & & 2013 Mar ­5 13:00& 2013 Mar ­6 13:00& 4867\\
  & & & 2013 Oct 18 00:00& 2013 Oct 19 00:00& 5506\\
\hline
  AGL J1647+5107  & Unidentified  &  1  & 2013 Feb ­22 15:00& 2013 Feb ­24 17:18 & 4842\\
\hline
 AGL J2302-3251  & Unidentified  &  1  & 2011 May ­14 03:00& 2011 May 17 03:00& 3357\\
\hline
  AGL J2241+4454  & Unidentified  &  1  & 2010 Jul ­25 01:00& 2010 Jul ­26 23:30&2761\\
\hline
  AGL J0906-1241  & Unidentified  &  1  & 2010 Apr ­11 05:00& 2010 Apr 13 06:00& 2552\\
\hline
  AGL J0109+6134  & Unidentified  &  1  & 2010 Jan ­31 08:20& 2010 Feb 2 19:48& 2416\\
\hline
  AGL J2206+6203  & Unidentified  &  1  & 2010 Jan ­20 03:46& 2010 Jan ­25 11:15& 2403\\
\hline
  AGL J1023-3738  & Unidentified  &  1  & 2009 Dec 25 22:30& 2009 Dec ­27 22:10& 2361\\
\hline
  MAXI J1659-152  & Unidentified  &  1  & 2010 Sep 25 00:00& 2010 Sep ­26 14:00& 2880\\
\hline
  AGL J0813+2420  & Unidentified  &  1  & 2010 Oct ­23 18:00& 2010 Oct ­26 06:00&2971\\
\hline
  AGL J1037-5708  & Unidentified  &  1  & 2010 Nov 27 21:18 & 2010 Nov ­30 14:08& 3059\\
\hline
  AGL J2103+5630  & Unidentified  &  1  & 2011 Aug 8 04:10& 2011 Aug ­10 09:04& 3544\\
\hline
  AGL J1524+3642  & Unidentified  &  1  & 2012 Jan ­9 11:26& 2012 Jan ­11 10:57& 3862\\
 \hline
\enddata
\label{tableA2}
\end{deluxetable}

\clearpage

\begin{figure}[!htb]
\centering
\includegraphics[width=18 cm]{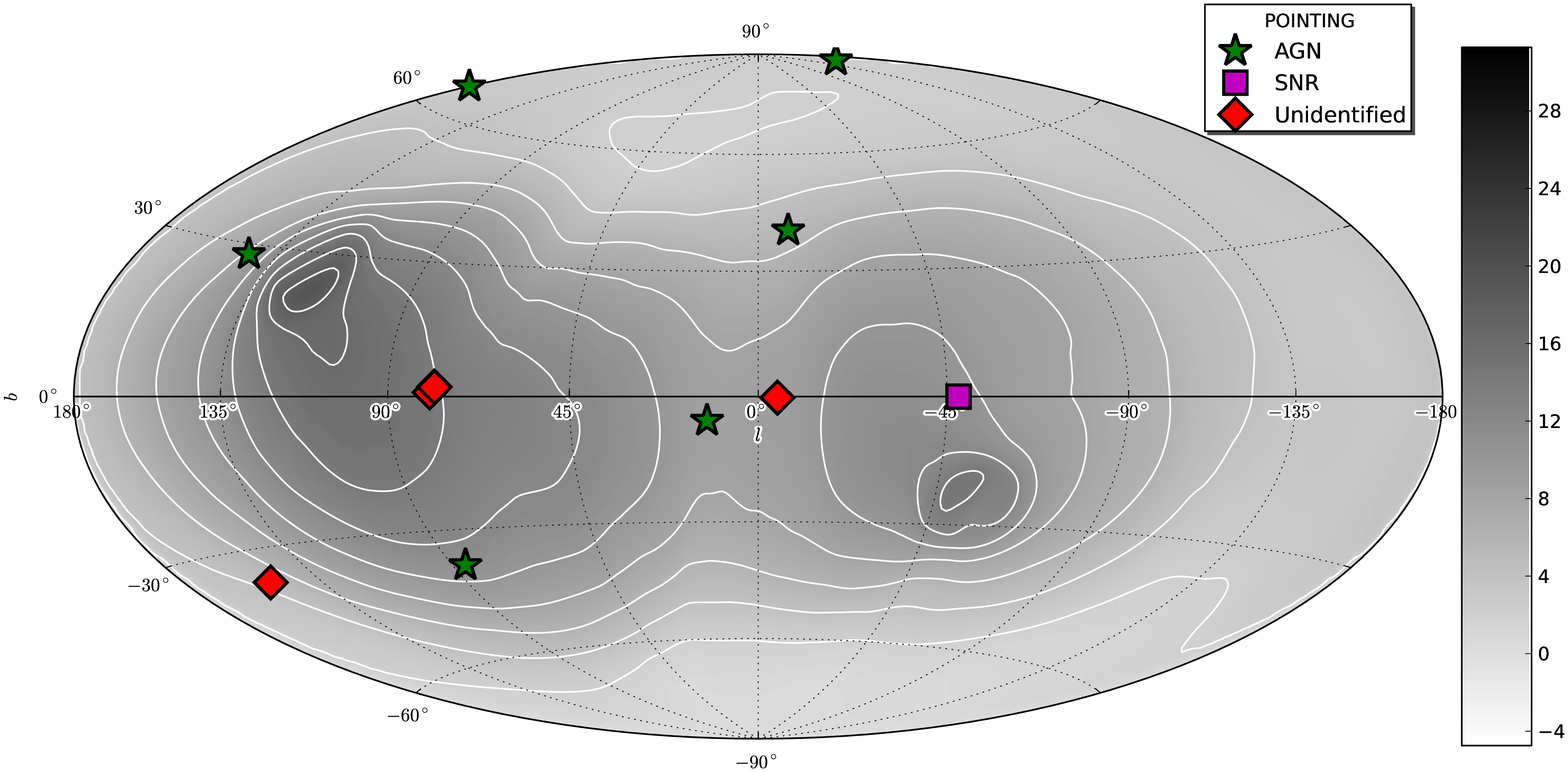}
\centering \caption { Position and classification of the published \textit{AGILE}-GRID Astronomical Telegrams in ``pointing'' mode overlapped to the exposure map.}
\label{fig_atel1}
\end{figure}

\begin{figure}[!htb]
\centering
\includegraphics[width=18 cm]{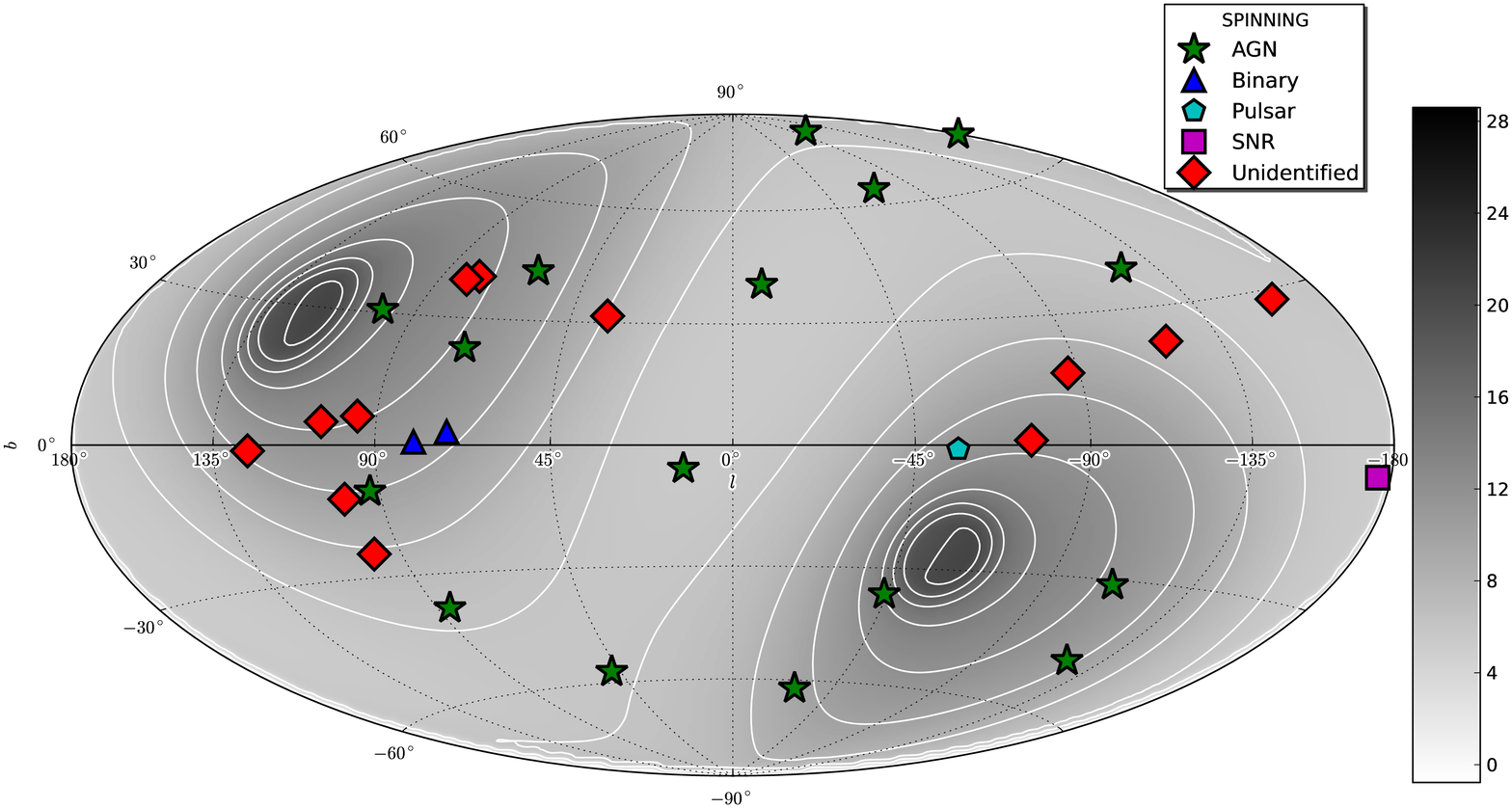}
\centering \caption { Position and classification of the published \textit{AGILE}-GRID Astronomical Telegrams in ``spinning'' mode overlapped to the exposure map.}
\label{fig_atel2}
\end{figure}

\end{document}